\documentclass[slac_one]{revtex4}
\usepackage{graphicx}
\usepackage{fancyhdr}
\begin{document}

\title{Searches for Physics Beyond the Standard Model at CMS} 

%

\author{Sung-Won Lee, on behalf of the CMS Collaboration}
\affiliation{Texas Tech University, Lubbock, TX 79409, USA}

\begin{abstract}
Recent results on searches for physics beyond the Standard Model at Large Hadron Collider are presented, based on 
early LHC data in proton-proton collisions at $\sqrt{s} = 7$ TeV collected by the CMS experiment. 
Prospects of early SUSY searches at CMS are also outlined.  

\end{abstract}

\maketitle

\thispagestyle{fancy}



\section{INTRODUCTION}
The Standard Model~(SM) has been enormously successful, but it leaves many
important questions unanswered. It is also widely acknowledged that, from the 
theoretical standpoint, the SM must be part of a larger theory,~``beyond" the SM~(BSM), 
which is yet to be experimentally confirmed.  

One of the most popular suggestions for the BSM theory is Supersymmetry (SUSY) which 
introduces a new symmetry between fundamental particles. SUSY signals are of particular interest, as they provide 
a natural explanation for the ÒDark MatterÓ, known to pervade our universe, and help us to 
understand the fundamental connection between particle physics and cosmology. Furthermore 
there are a large number of important and well thought out theoretical models that make strong 
cases for looking for new physics at the LHC. These theories include Extra 
Dimensions, Black Hole, Grand Unified Theories, Composite models, Anomalous couplings and 
non-SM Higgs models. None of the rich new spectrum of particles predicted by these models have yet been 
found within the kinematic regime reachable at the present experiments. The LHC will increase 
this range dramatically after several years of running at the highest energy and luminosity. 

One of the primary objectives of CMS experiments is  to find incontrovertible evidence for new 
physics beyond SM using a signature of high-energy objects in the final state, and the `signatures' 
expected for new physics have been taken into consideration extensively in the design of the experiment~\cite{CMS}. 
In this article we summarize the current experimental results of 
searches for physics beyond the SM from the CMS experiment at the Large Hadron Collider~(LHC). 


\section{SEARCHES FOR DIJET RESONANCES}

The signatures of new physics at the LHC involve high-$p_T$ final-state jets. 
The dijet mass spectrum predicted by  Quantum Chromodynamics (QCD) falls smoothly and steeply 
with increasing dijet mass. As an example of the new physics, many extensions of the SM predict 
the existence of new massive objects that couple to quarks and gluons, and result in resonant structures 
in the dijet mass spectrum.  CMS has performed a search for narrow resonances in the dijet mass spectrum 
using 2.9 pb$^{-1}$ of early LHC data, at a proton-proton collision energy of $\sqrt{s} = 7$ TeV ~\cite{CMS_Dijetmass}. 

The dijet system is composed of the two jets with the highest $p_T$ in an event (leading jets). 
CMS requires that the pseudorapidity separation of the two leading jets, $\Delta\eta=\eta_1-\eta_2$, 
satisfies $|\Delta\eta|<1.3$, and that both jets be in the region $|\eta|<2.5$. 
These $\eta$ cuts maximize the search sensitivity for isotropic decays of dijet resonances 
in the presence of QCD background. The dijet mass is given by 
$m=\sqrt{(E_1 + E_2)^2 - (\vec{p}_1 + \vec{p}_2)^2}$. CMS selects events with $m>220$~GeV 
without any requirements on jet $p_T$.

\begin{figure*}[t]
\centering
\includegraphics[width=92mm]{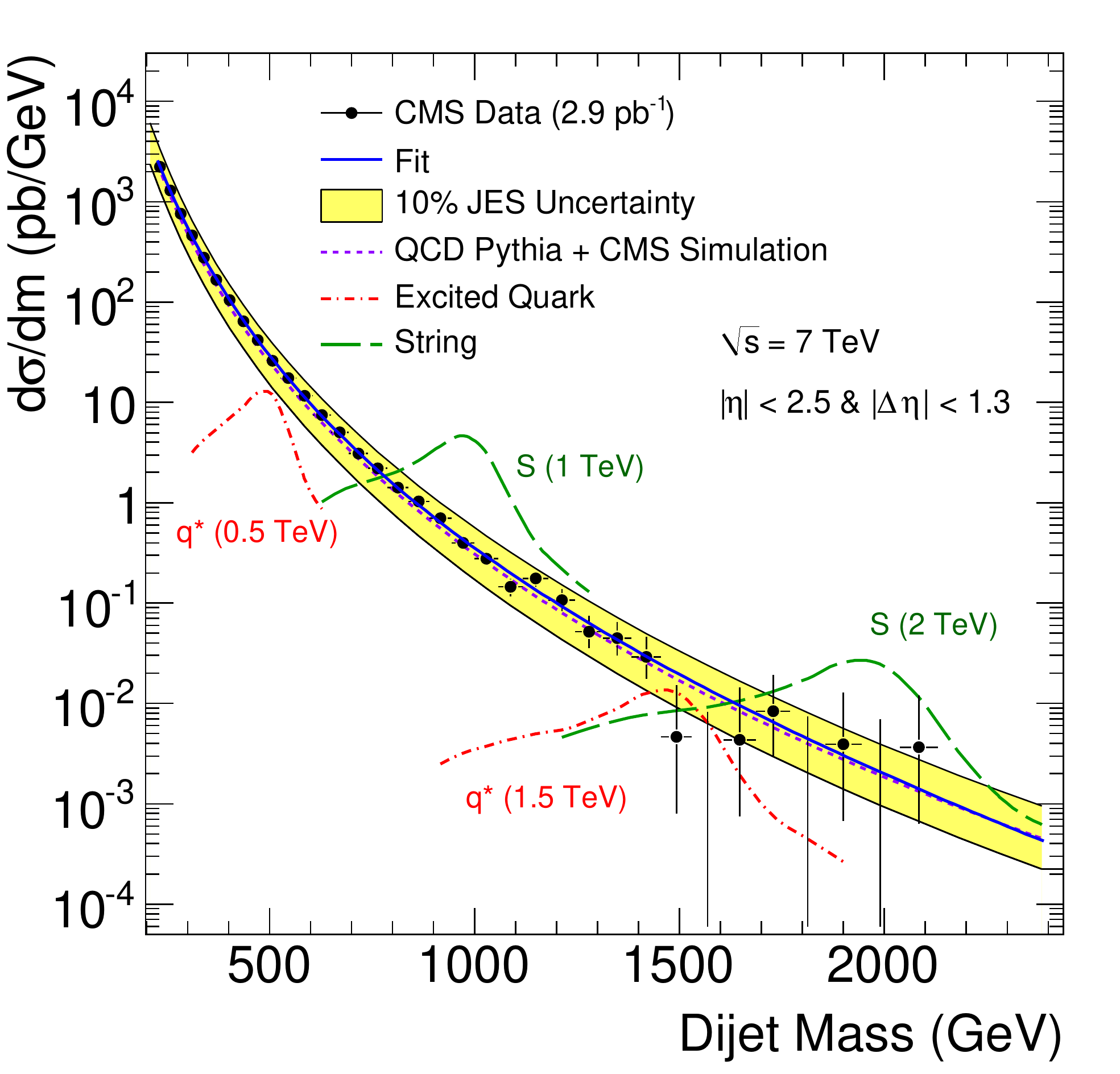}
\includegraphics[width=84mm]{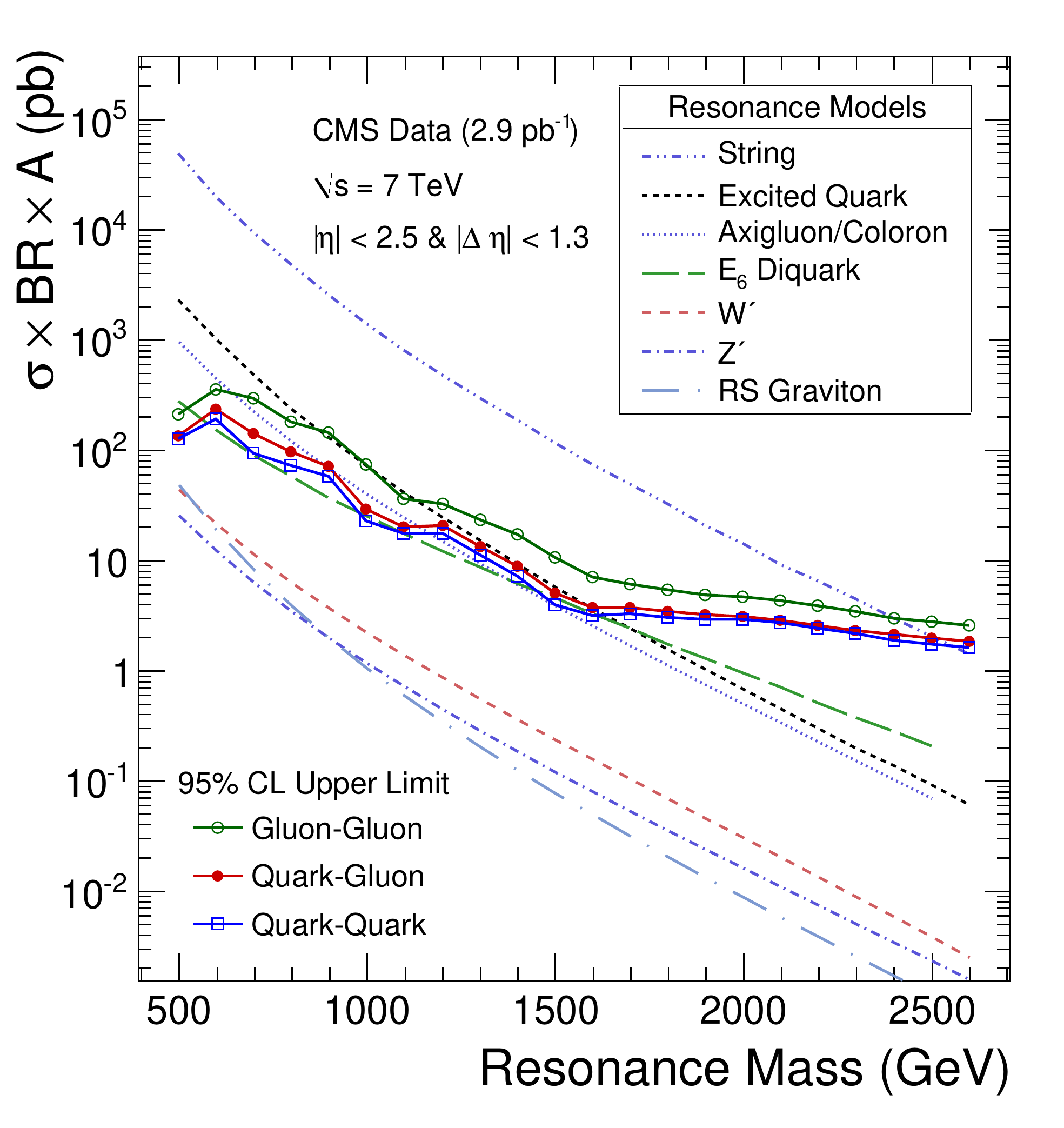}
\caption{(Left)~Dijet mass spectrum (\textit{points}) compared to a smooth fit (\textit{solid}) 
and to PYTHIA predictions including detector simulation of QCD (\textit{short-dashed}), 
excited quark signals (\textit{dot-dashed}), and string resonance signals (\textit{long-dashed}). The errors are statistical only. 
The shaded band shows the effect of a 10\% systematic uncertainty in the jet energy scale. 
(Right)~95\% CL upper limits on $\sigma\times\mbox{BR}\times\mbox{A}$ 
for dijet resonances of type gluon-gluon (\textit{open circles}), quark-gluon (\textit{solid circles}), 
and quark-quark (\textit{open boxes}), 
compared to theoretical predictions for string resonances, excited quarks, axigluons, 
colorons, $\mbox{E}_6$ diquarks, new gauge bosons $W^{\prime}$ and $Z^{\prime}$,
and Randall-Sundrum gravitons.} 
\label{dijet_mass}
\end{figure*}

Figure~\ref{dijet_mass}~(left) presents the inclusive dijet mass distribution for 
$pp\rightarrow$ 2 leading jets + $X$, where $X$ can be anything, including additional jets. 
The data are compared to a QCD prediction from PYTHIA, which includes a simulation of the CMS  
detector and the jet energy corrections~\cite{CMS_JEC}.  The PYTHIA prediction agrees with the data within 
the jet energy scale uncertainty, which is the dominant systematic uncertainty. 
The measured dijet mass spectrum is a smoothly falling distribution as expected within the SM and 
there is no indication of narrow resonances in the data. 
Figure~\ref{dijet_mass}~(left)  also shows the predicted dijet mass distribution for string resonances 
and excited quarks models. The predicted mass distribution exhibits  a Gaussian core from 
jet energy resolution and a tail towards low mass from QCD radiation. 

Since no significant excess over the SM prediction is observed,  
CMS presents  generic upper limits at the 95\% C.L. on the product of the resonance cross 
section, branching fraction into dijets, and acceptance, as a function of the new particle mass, 
for narrow resonances decaying to dijets with partons of type quark-quark ($qq$), quark-gluon ($qg$), 
and gluon-gluon ($gg$)~\cite{CMS_search} .  These generic limits are used to exclude new particles predicted in the following 
specific models:   string resonances~(S), excited quarks~(q$^{*}$), axigluons~(A), flavor universal colorons~(C),  
and $\mbox{E}_6$ diquarks~(D). 

In Figure~\ref{dijet_mass}~(right)  CMS compares these upper limits to the model predictions 
as a function of resonance mass. CMS excludes at the 95\% C.L. new particles in mass regions for 
which the theory curve lies above upper limit for the appropriate pair of
partons.  For string resonances CMS uses the limits on $qg$ resonances to 
exclude the mass range $0.50 < M(S) < 2.50$~TeV.
For comparison, previous measurements~\cite{refCDFrun2} 
imply a limit on string resonances of about 1.4~TeV.
For excited quarks CMS excludes the mass range $0.50<M(q^*)<1.58$~TeV,
extending the previous exclusion of $0.40<M(q^*)<1.26$~TeV~\cite{ATLAS_search}. 
For axigluons or colorons CMS uses the limits on $qq$ resonances to exclude the mass intervals 
$0.50<M(A)<1.17$~TeV and $1.47<M(A)<1.52$~TeV, extending the
previous exclusion of $0.12<M(A)<1.25$~TeV~\cite{refCDFrun2}. For $\mbox{E}_6$ diquarks CMS excludes the mass
intervals $0.50<M(D)<0.58$~TeV, and $0.97 < M(D) < 1.08$~TeV, and $1.45 < M(D) < 1.60$~TeV, 
extending the previous exclusion of $0.29<M(D)<0.63$~TeV~\cite{refCDFrun2}.
For $W^\prime$, $Z^\prime$ and RS gravitons CMS does not expect any mass limit, and does not exclude any mass intervals 
with the present data.
The systematic uncertainties included in this analysis reduce the excluded upper masses 
by roughly $0.1$~TeV for each type of new particle.


\section{SEARCHES FOR QUARK COMPOSITENESS}

In the SM production of dijet events, the pseudorapidity $\eta$ of the jets depends 
on the angular distribution of the scattered partons predicted by QCD. 
New physics beyond the SM, including models of quark
compositeness, typically produces more isotropic angular distributions than those 
predicted by QCD, resulting in more dijets at lower absolute values of pseudorapidity. 

CMS has searched for quark compositeness in the framework of quark contact
interactions, based on the dijet centrality ratio, $R_{\eta} = N(|\eta|<0.7)$ / $N(0.7<|\eta|<1.3)$, 
which is the number of events with the two leading jets in the region $|\eta|<0.7$ (inner
events) divided by the number of events with the two jets in the region
$0.7<|\eta|<1.3$ (outer events), using a data sample corresponding to 120 $ \pm $ 13 nb$^{-1}$ 
of integrated luminosity~\cite{CMS_Dijetratio}. Since many sources of systematic uncertainty cancel in this ratio, 
the dijet ratio provides a precise test of QCD and is sensitive to new physics. 
This analysis is closely related to the CMS search for dijet resonances in the dijet mass 
spectrum, described in Section 2. Though the centrality ratio analysis is less sensitive in the 
case of resonances, it is more sensitive to the presence of contact interactions.

\begin{figure*}[t]
\centering
\includegraphics[width=88mm]{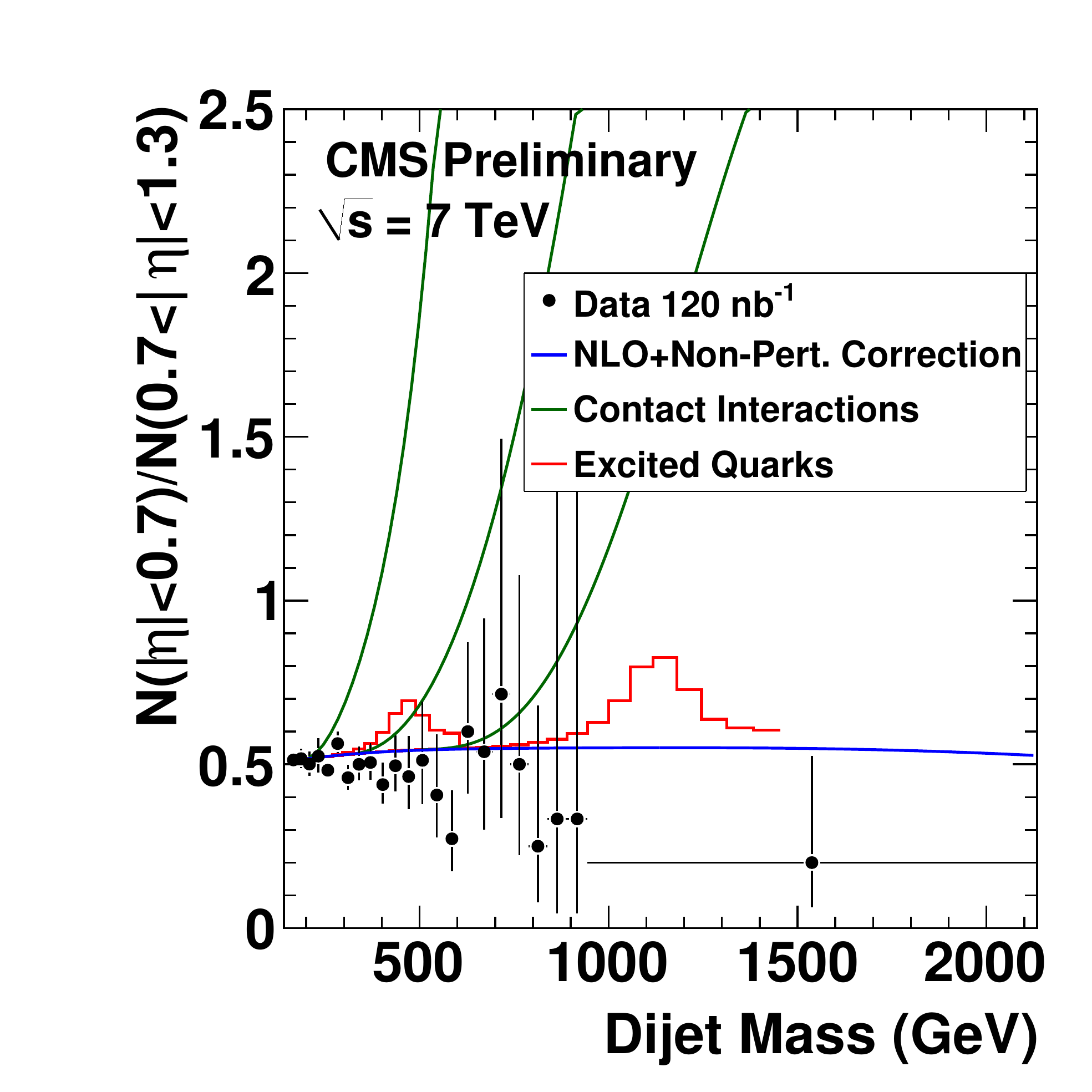}
\includegraphics[width=88mm]{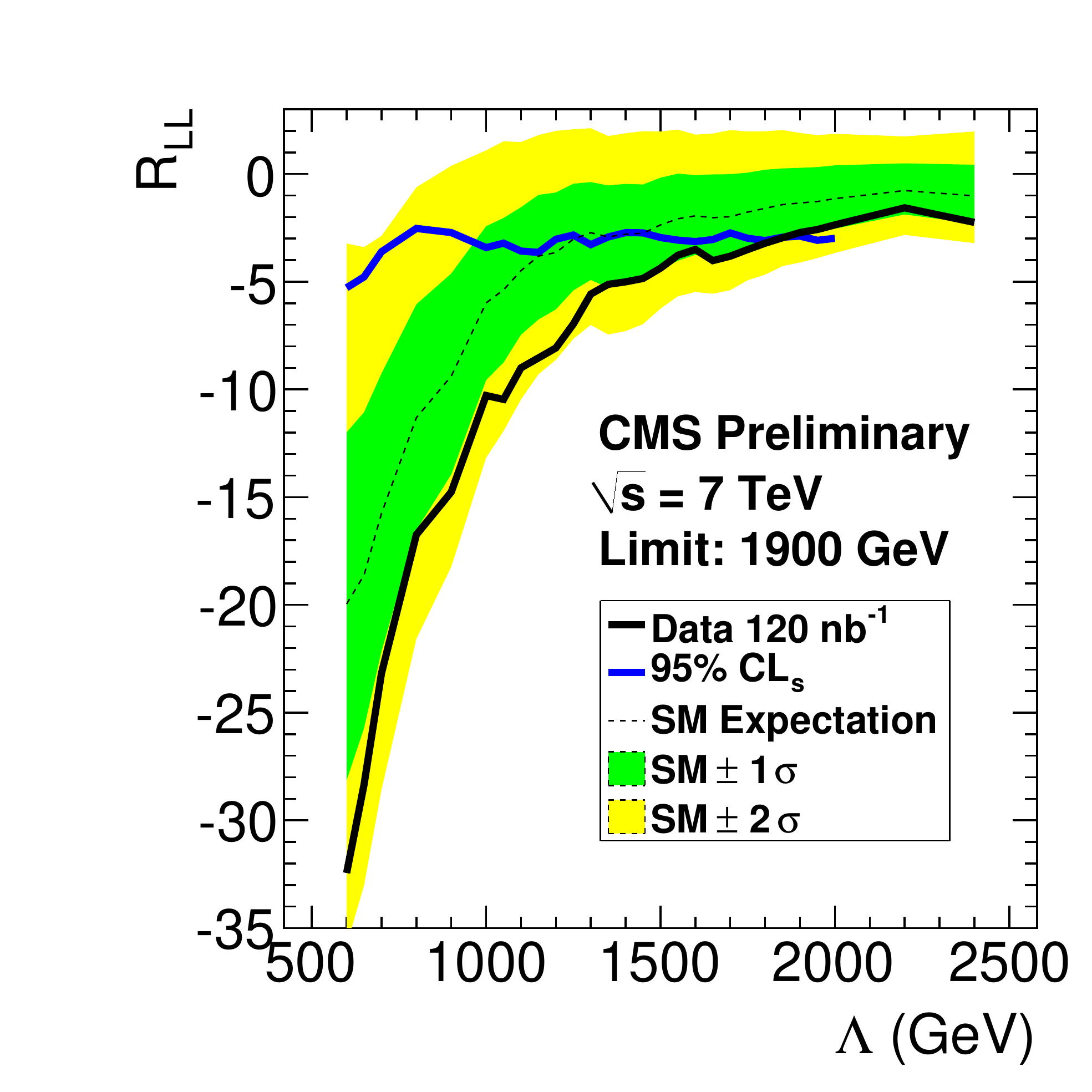}
\caption{
(Left)~Dijet centrality ratio in 120 nb$^{-1}$ of integrated luminosity. 
The ratio is compared to predictions for NLO QCD plus non-perturbative corrections, 
contact interactions with $\Lambda$ = 0.5, 1.0, and 1.5 TeV, and excited quark 
resonances with masses of 0.5 and 1.2 TeV. 
(Right)~Summary of the limit for the contact interaction scale $\Lambda$.  
We show ${\cal R}_{LL}$ versus $\Lambda$ for the data 
(\textit{solid black line}), the 95\% CL$_s$ (\textit{solid blue line}), 
and the SM expectation (\textit{dashed black line}) with 
1$\sigma$ (\textit{green}) and 2$\sigma$ (\textit{yellow}) bands.
} 
\label{dijet_ratio}
\end{figure*}

Figure~\ref{dijet_ratio}~(left) shows a comparison of the measured dijet centrality ratio with 
the predictions of NLO QCD and various new physics models. The dijet centrality ratio 
 in data is nearly flat as predicted by QCD. To quantitatively test for 
the presence of new physics in the dijet centrality ratio, CMS uses a log-likelihood-ratio 
statistic~(${\cal R}_{LL}$) that compares the null hypothesis (SM only) to the hypothesis 
that new physics effects are present in addition to the SM.  Given this consistency of the data 
with the QCD hypothesis, CMS has determined 95\% CL limits on the contact interaction scale
$\Lambda$.
Figure~\ref{dijet_ratio}~(right) shows ${\cal R}_{LL}$ versus $\Lambda$ for the data and 
for the SM expectation (with 1$\sigma$ and 2$\sigma$ bands) along with the highest value 
of ${\cal R}_{LL}$ excluded at the 95\% CL with the CL$_s$ method~\cite{CMS_Dijetratio2}.
CMS excludes quark compositeness described
by a contact interaction between left-handed quark fields at energy scales
of $\Lambda<1.9$ TeV at the 95\% C.L.


\section{SEARCH FOR STOPPED GLUINOS}

CMS has searched for long-lived gluinos which have stopped in the CMS detector 
after being produced in 7 TeV $pp$ collisions from LHC~\cite{CMS_SG}.  
CMS looked for the subsequent decay of these particles during time intervals where 
there were no $pp$ collisions in the CMS experiment.  
In particular, CMS has searched for decays during gaps between crossings 
in the LHC beam structure, and recorded such decays with a dedicated calorimeter trigger.  
In a dataset  with a peak instantaneous luminosity of $1.3 \times 10^{30} {\rm cm}^{-2} {\rm s}^{-1}$, 
an integrated luminosity of 203 - 232 nb$^{-1}$, depending on the gluino lifetime, and a search 
interval corresponding to 115 hours of LHC operation, no significant excess above background 
was observed.  In the absence of a signal, CMS set a limit at 95\%~C.L. on gluino pair production 
over 14 orders of magnitude of gluino lifetime. 
For a mass difference $m_{\tilde{g}}-M_{\tilde{\chi}^0_1} >100$ GeV, assuming 
BR($\tilde{g} \rightarrow g\tilde{\chi}^0_1$) = 100\%, CMS excludes lifetimes from 75 ns 
to 6 $\mu$s for $m_{\tilde{g}} = 200$ GeV/$c^2$.  

\begin{figure*}[t]
\centering
\includegraphics[width=88mm]{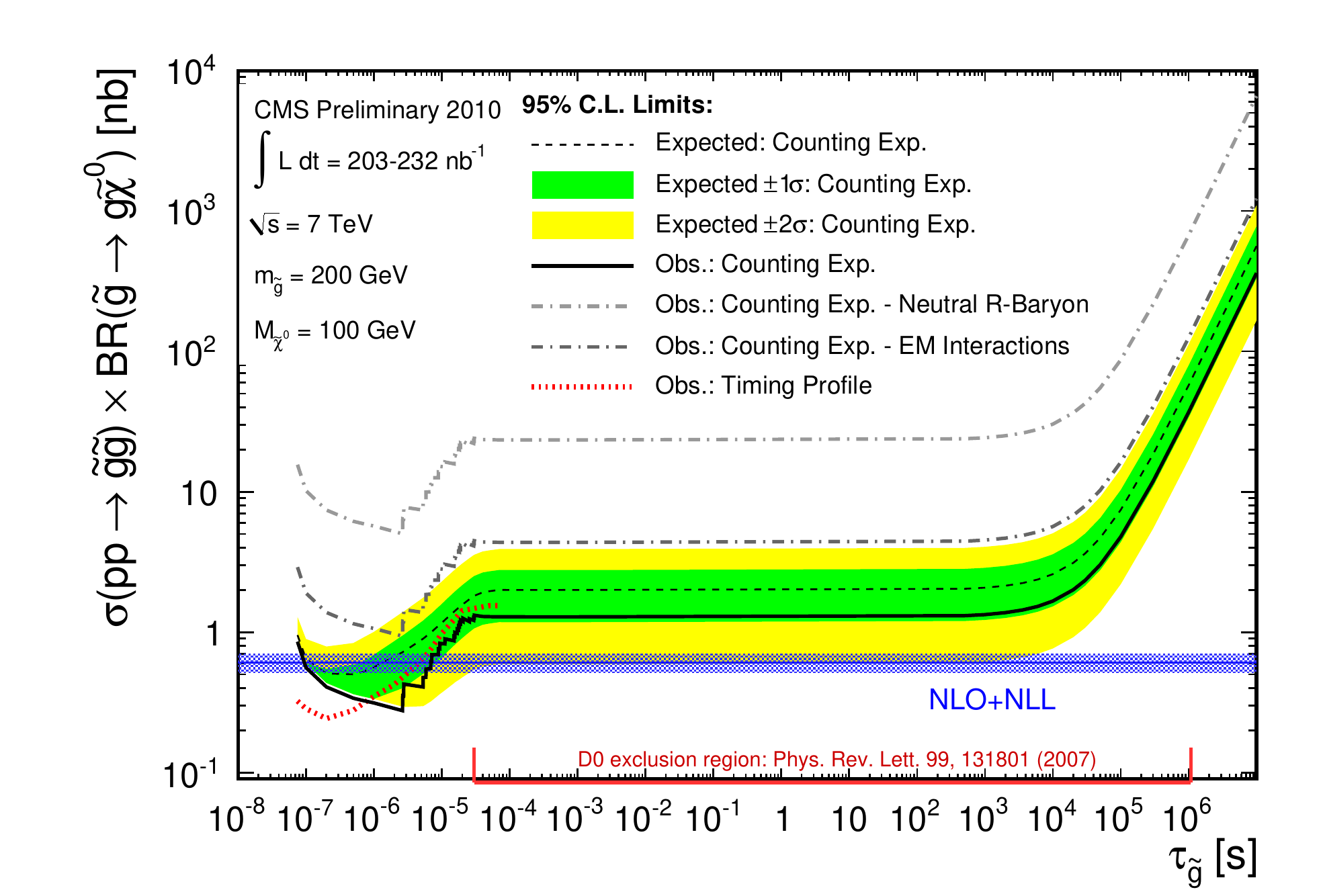}
\includegraphics[width=88mm]{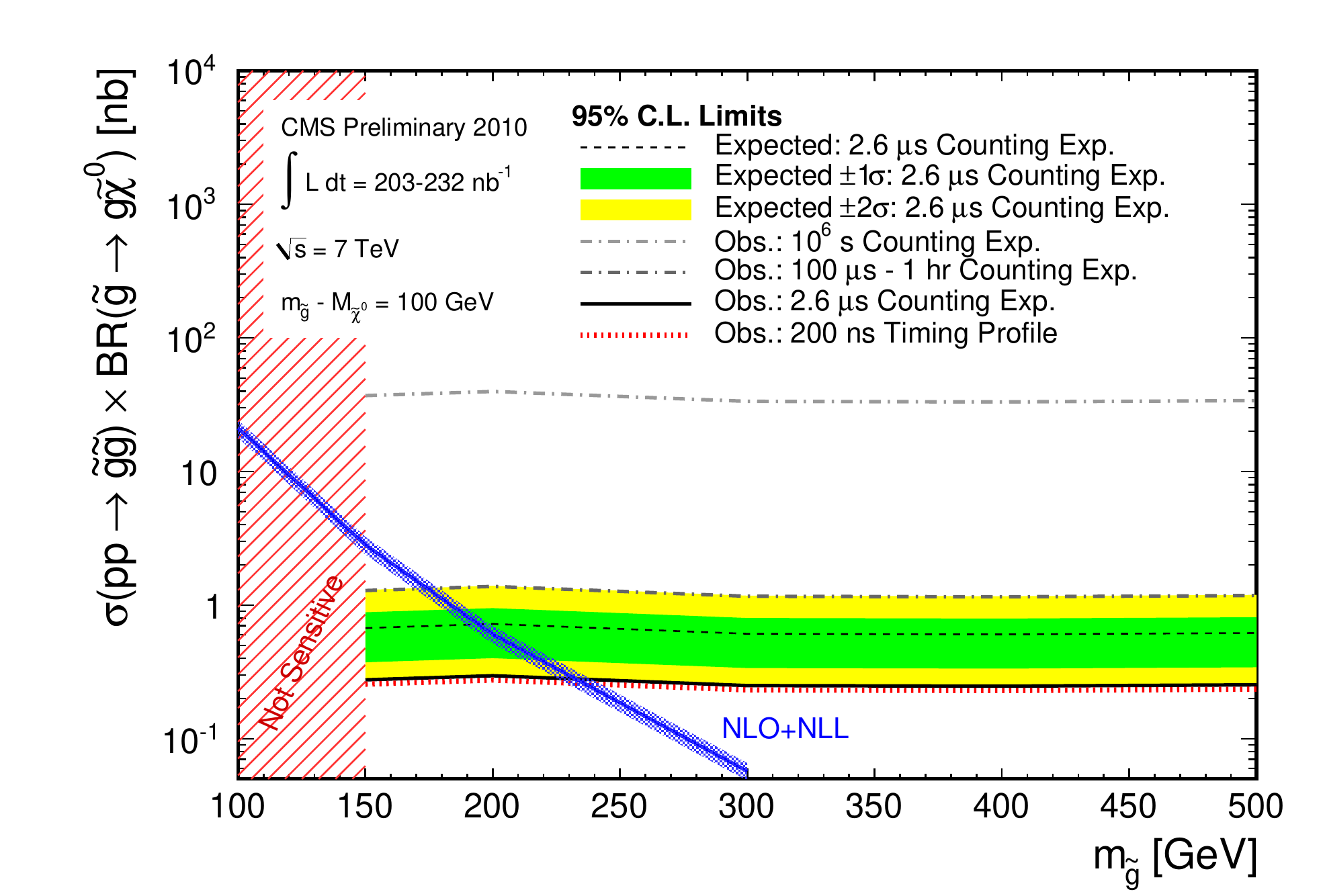}
\caption{
(Left)~Expected and observed 95\% C.L. limits on gluino pair production cross-section using the ``cloud model'' of R-hadron interactions as a function of gluino lifetime from both counting experiment and the time profile analysis. Observed 95\% C.L. limits on the gluino cross-section for alternative R-hadron interaction models are also presented. 
(Right)~95\% C.L. limits on gluino pair production cross-section as a function of gluino mass assuming the ``cloud model'' of R-hadron interactions.  The $m_{\tilde{g}}-M_{\tilde{\chi}^0_1}$ mass difference is maintained at 100 GeV; results are only presented for $M_{\tilde{\chi}^0_1} > 50$ GeV.   
} 
\label{fig1}
\end{figure*}

These results extend existing limits from the Tevatron, which exclude lifetimes between 30 $\mu$s and 100 hours~\cite{CMS_SG1}. 
 Furthermore CMS excludes gluino masses $m_{\tilde{g}} < 229$ GeV/$c^2$ with a lifetime of 200~ns using the time-profile analysis and $m_{\tilde{g}} < 225$ GeV/$c^2$ with a lifetime of 2.6~$\mu$s in a counting experiment.  This result is consistent with the complementary exclusion provided by our direct HSCP search~\cite{CMS_HSCP}.  As more luminosity is delivered by the LHC the reach of this analysis will improve rapidly.  In particular, since the only backgrounds to this search are independent of luminosity, this sensitivity will increase significantly when the LHC peak instantaneous luminosity increases to $10^{32}$~cm$^{-2}$s$^{-1}$ expected later this year.


\section{SEARCH FOR HEAVY STABLE CHARGED PARTICLES}

Heavy Stable (or long-lived) Charged Particles (HSCPs) appear in various extensions 
to the SM, arising from a new symmetry,  a weak coupling, a kinematic 
constraint, or a potential barrier. 
If the lifetime is long compared to the transit time through the detector, 
then the particle may escape the detector, thereby
evading the limits imposed by direct searches for decay products. 
Nevertheless, a HSCP will be directly observable in 
the detector through the distinctive signature of a slowly moving,
high  momentum ($p$) particle. The low velocity results in 
an anomalously large ionization-energy loss rate ($dE/dx$).

\begin{figure*}[t]
\centering
\includegraphics[width=88mm]{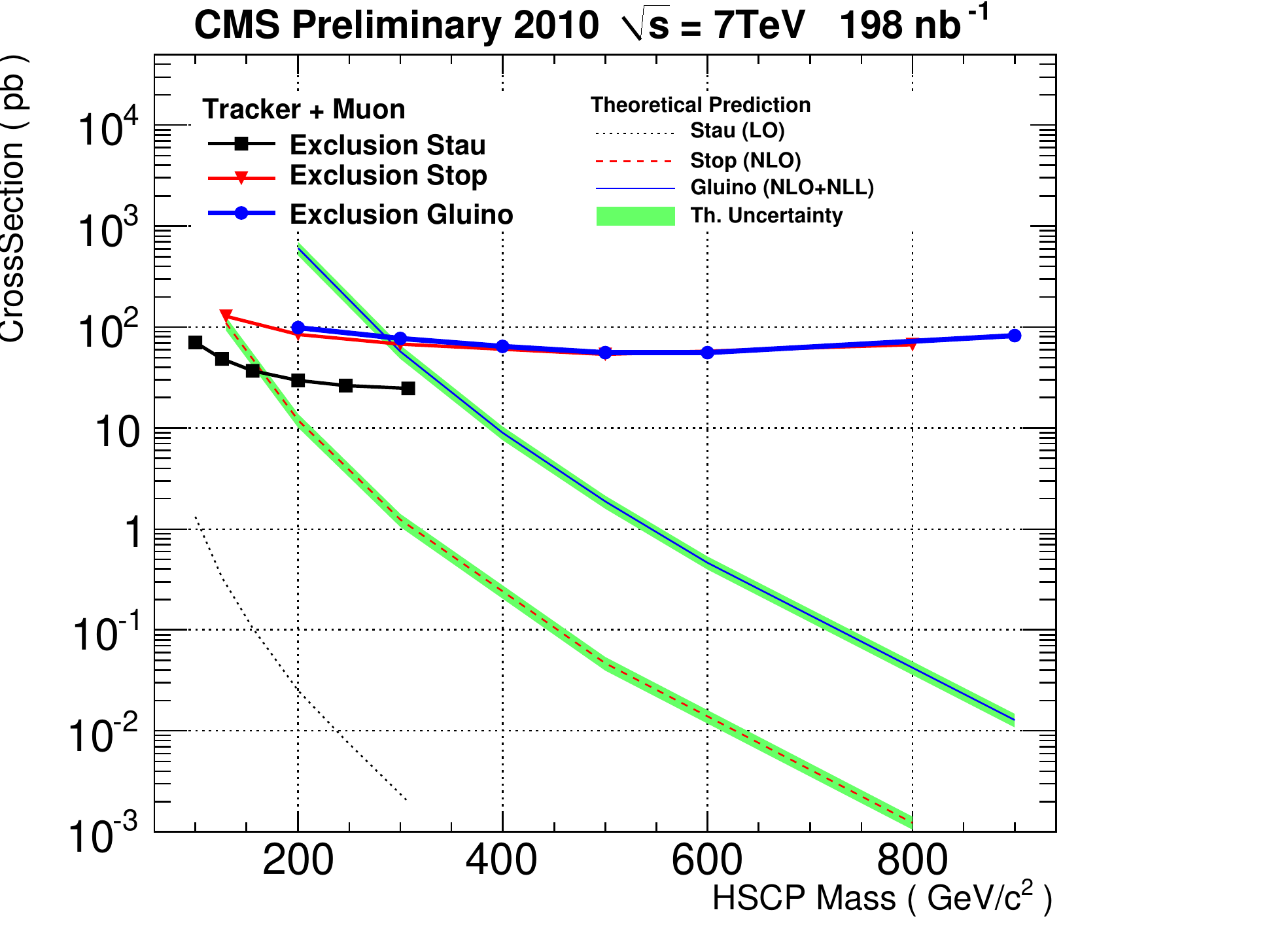}
\includegraphics[width=88mm]{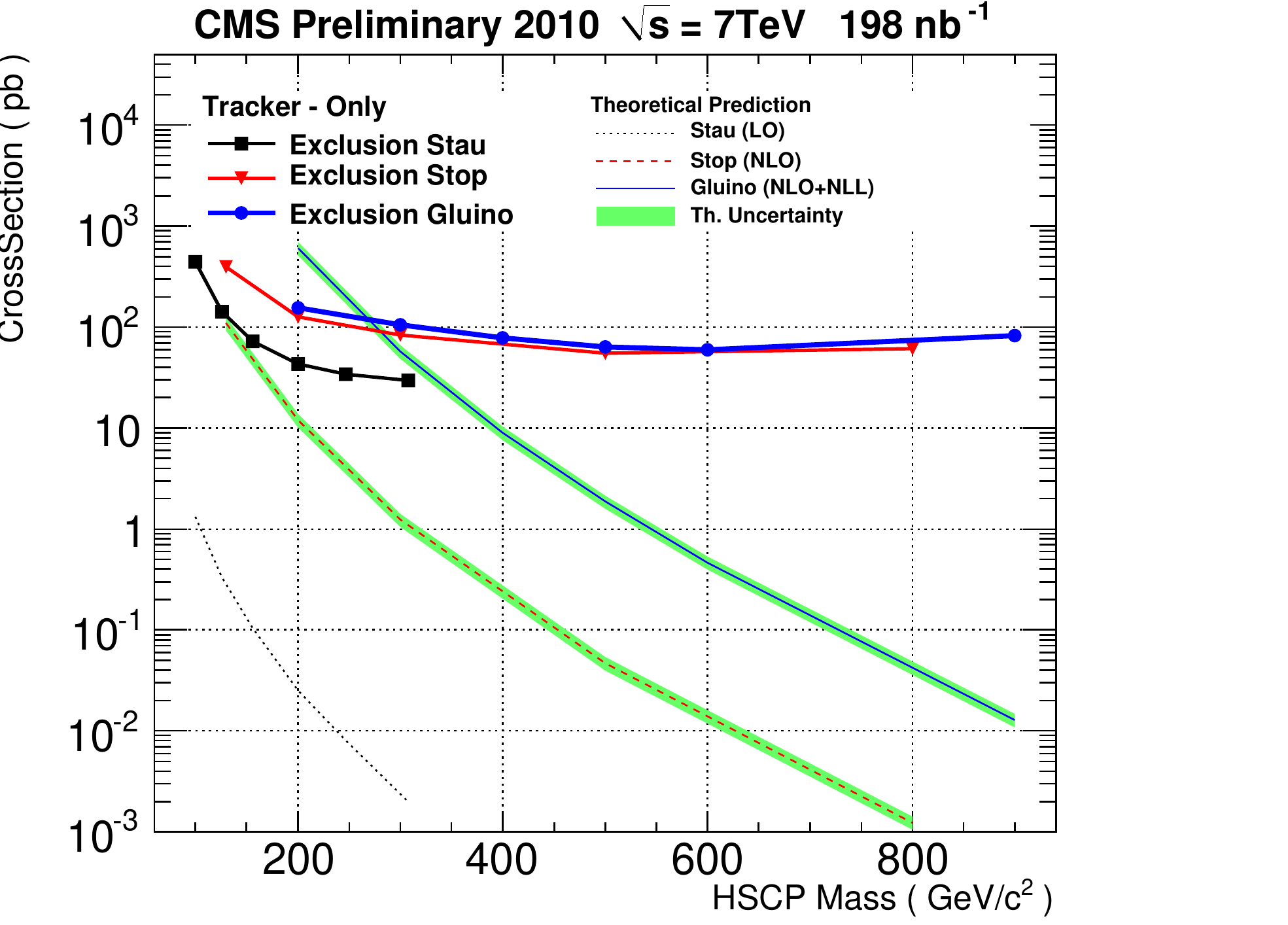}
\caption{
Observed 95\% C.L. upper limits on the cross section for production 
 of the different models considered and predicted theoretical cross sections.  
(Left)~Analysis of the muon identification plus tracker candidates; 
(Right)~Analysis of the tracker-only candidates. 
The bands represents the theoretical uncertainty on the cross section values.    
} 
\label{fig1}
\end{figure*}

In this analysis, a signature-based search is performed for HSCPs produced 
in $pp$ collisions at $\sqrt{s} = 7$ TeV, 
using high transverse momentum muon, jet, and missing transverse energy~(MET) trigger 
data corresponding to 198 nb$^{-1}$ of integrated luminosity~\cite{CMS_HSCP}.  
The analysis isolates HSCP candidates by selecting 
tracks reconstructed in the inner tracker detector 
with high $dE/dx$ and high $p_{T}$. 
 Additionally, tracks passing muon
identification requirements are also analyzed for this signature.  
For both selections, the candidate's mass is then calculated from
the measured $p$ and $dE/dx$.  
In both cases, no event passes the selection criteria with an expected 
background of less than 0.1 events. 
From this result, an upper limit at 95\% C.L. 
on the production cross section of 
pairs of stable gluinos, hadronizing into R-gluonballs
in 10\% of the cases, and top squarks is set at around 10 pb
starting from a mass of 130 and 200 GeV/$c^2$, respectively. 
For the case of gluinos a mass lower limit of 382 (190) GeV/$c^2$  
can be set at 95\% C.L. with the analysis that uses muon
identification. This limit becomes 375 (183) GeV/$c^2$ when no muon
identification is required. 
Cross section upper limits are also set for some benchmark points 
in the framework of the mGMSB model, predicting the existence of
stable staus.


\section{EARLY LHC DATA PREPARATION FOR SUSY SEARCHES}
CMS will perform a broad range of searches for SUSY particles. The initial searches will be performed 
in a variety of inclusive final states involving jets, leptons, photons, and MET. 
These searches require careful control over backgrounds from SM processes. Several methods for data-driven background determinations were developed and tested on early LHC data collected by CMS experiment~\cite{CMS_SUSY1}. 
These data allow us to study QCD backgrounds, to evaluate methods to suppress the effects of jet-energy 
mismeasurement, to validate data-driven methods for predicting the background MET distribution, and to measure background contributions from processes producing non-prompt leptons or hadrons misidentified as leptons. 

\begin{figure*}[t]
\centering
\includegraphics[width=112mm]{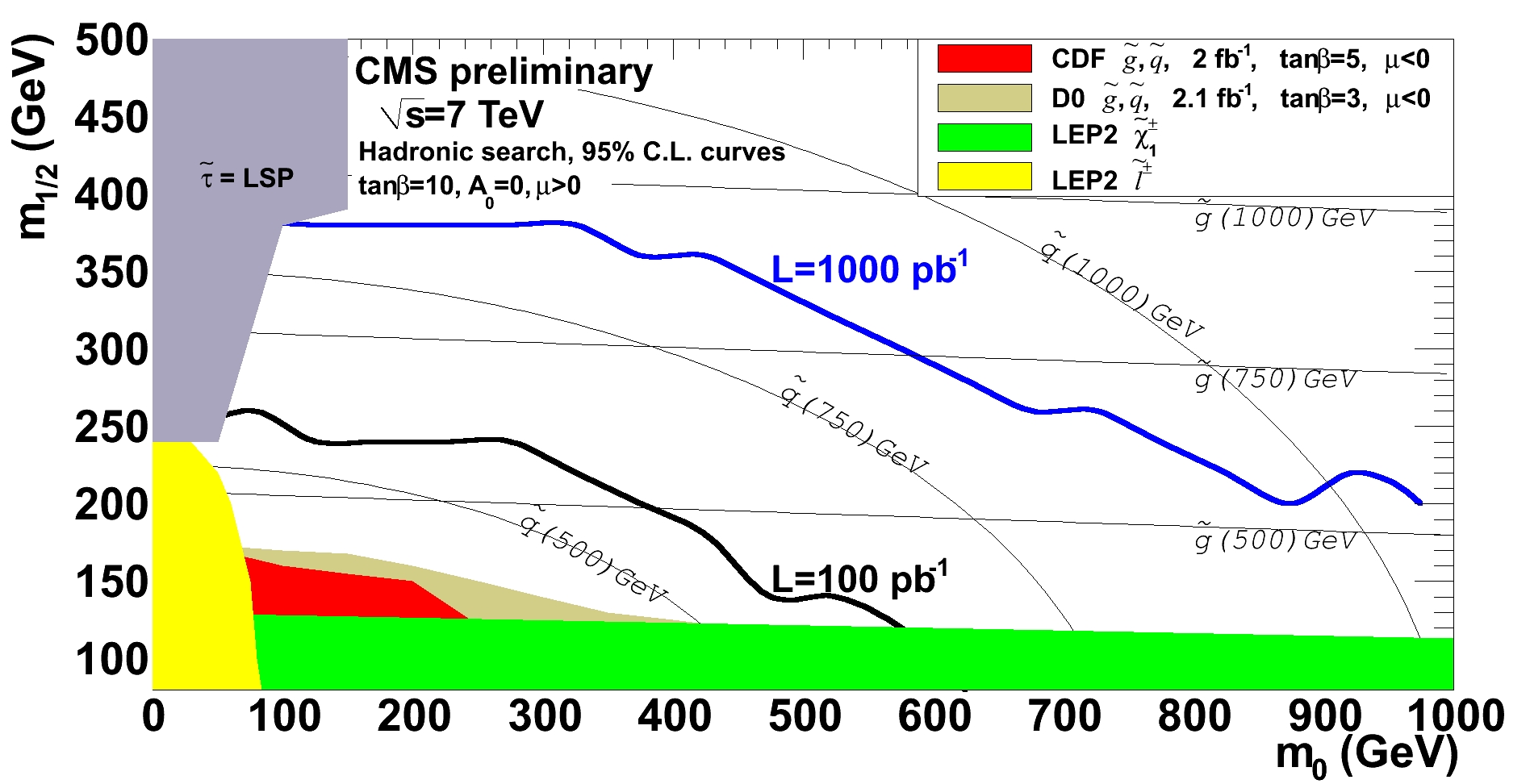}
\includegraphics[width=58mm]{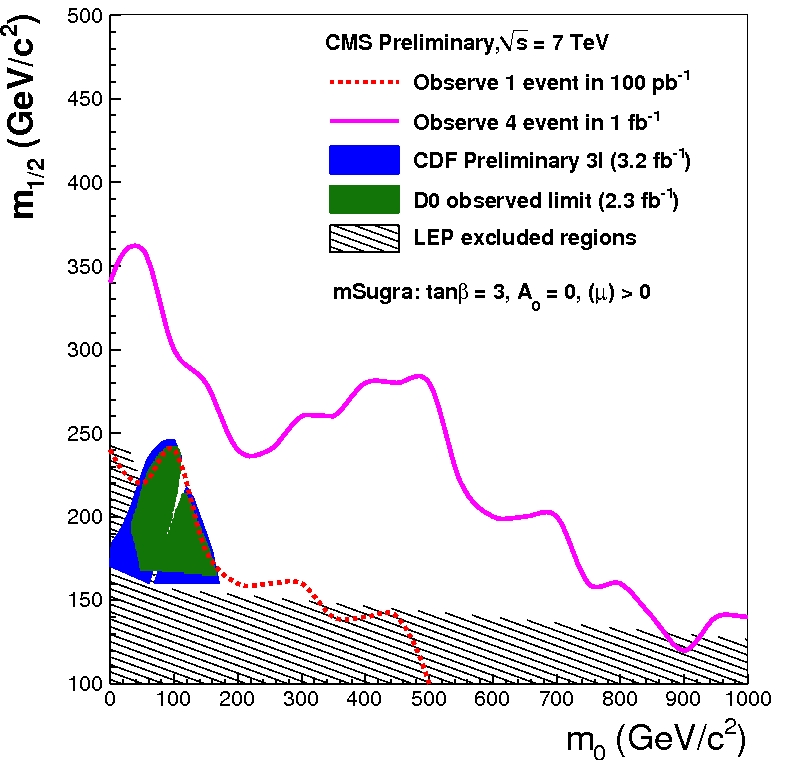}
\caption{
(Left)~Estimated 95\% C.L. exclusion limits for the all-hadronic SUSY search, expressed in mSUGRA parameter space. 
(Right)~Estimated 95\% C.L.  exclusion limits for the like-sign dilepton SUSY search, expressed in mSUGRA parameter space. 
The expected SM background at 100 pb$^{-1}$ (1 fb$^{-1}$) is 0.4 (4.0) events; an observed yield of 
1 event (4 events) is assumed for the purpose of setting these exclusion limits.
} 
\label{SUSY}
\end{figure*}

The prospects for the discovery reach are studied with two values of the integrated luminosity, 100 pb$^{-1}$  and 1 fb$^{-1}$ 
of simulated data at  $\sqrt{s} = 7$ TeV 
and are shown in the mSUGRA model in Figure~\ref{SUSY} for the all-hadronic channel (left) and 
for the like-sign dilepton channel (right)~\cite{CMS_SUSY2}. These results indicate that in the 7 TeV run, 
CMS should have sensitivity to regions of SUSY (mSUGRA) parameter space beyond the current Tevatron limits. 
Both of the channels shown here (all-hadronic and like-sign dileptons) should be able to yield interesting 
sensitivities well before 1  fb$^{-1}$.

\section{CONCLUSIONS}
The CMS experiment has searched for evidence of different models of new physics in several channels using early LHC data 
and already explored new territory beyond the Tevatron. 
No evidence for new physics signature has yet been observed in the early LHC data.  
The CMS collaboration expects to collect 1 fb$^{-1}$  of data by the end of 2011. 
This new data will make significant advances across a wide range 
of physics channels and will provide a great opportunity 
for new physics discovery.


\end{document}